\documentclass[final,3p,times,twocolumn]{elsarticle}

\usepackage{amsmath}
\usepackage{amssymb}
\usepackage{amsfonts}
\usepackage{mathtools}
\allowdisplaybreaks[1] 
\DeclareMathOperator{\tr}{tr}
\DeclareMathOperator{\grad}{grad}
\DeclareMathOperator{\divg}{div}

\DeclareMathAlphabet\mathbfcal{OMS}{cmsy}{b}{n}

\usepackage{graphicx}

\usepackage{hyperref}

\begin{document}
	

\title{Generalization of nonlinear Murnaghan elastic model for viscoelastic materials}

\author[1]{F.\,E. Garbuzov\corref{cor1}}
\ead{fedor.garbuzov@mail.ioffe.ru}
\cortext[cor1]{Corresponding author}

\author[1]{Y.\,M. Beltukov}

\address[1]{Ioffe Institute, 26 Polytekhnicheskaya, St.~Petersburg 194021, Russia}

\journal{International Journal of Non-Linear Mechanics}

\date{May 24, 2023}

\begin{abstract}
	This paper presents a generalization of Murnaghan elastic material to viscoelastic behavior using the Green-Rivlin multiple-integral approach. In the linear limit, the model coincides with the generalized Maxwell model. To create a nonlinear generalization, all possible second-order corrections were included in the constitutive equations written in the internal strains representation. Using this approach, we obtained expressions for the time- and frequency-dependent nonlinear dynamic moduli.
    We applied the developed nonlinear viscoelastic model to the description of infinitesimal strain waves superposed on finite prestrain. Furthermore, we considered the generation of higher harmonic by the nonlinear interaction of two strain waves, which we showed can provide a method to measure all viscoelastic constants of the developed model.
\end{abstract}

%


\maketitle

\section{Introduction}

The study of nonlinear materials, such as polymers, biological tissues, and geomaterials to name a few, is critical in various fields of science and engineering as they are ubiquitously present in the world.
Understanding the mechanical behavior of these materials can lead to the development of new materials with improved mechanical properties, the design of more efficient and reliable structures and devices, and the invention of new techniques for material quality assessment.

The most general model of an isotropic elastic material in the small but finite strain regime is the Murnaghan material, in which non-linearity appears as the next-order correction to the linear Hooke's law of elasticity~\cite{Murnaghan1951, LurieNonlin}. This model was used to obtain important results for the nonlinear strain waves.
In particular, the existence of bulk strain solitons in various thin structures was shown~\cite{SamsonovBook, Porubov2004} and the possible application of strain solitons to nondestructive testing and material properties inspection was extensively studied~\cite{KhusnSamsPhysRev2008, IJNM2017, Tranter2023}.

Besides the elastic response, many mechanically nonlinear materials exhibit viscoelastic behavior, as their mechanical properties are time and frequency dependent.
Models which combine nonlinear elasticity with linear viscosity have been successfully used in many important applications, such as the generation of a solitary strain wave~\cite{WaMot2022}, dispersive shock waves generated by fracture~\cite{KhusnPRE2021}, the split-Hopkinson bar technique~\cite{Wang1994}, and various nonlinear wave propagation problems~\cite{Destrade2013, Zabolotskaya2004}. However, the nonlinear viscous effects are rarely mentioned in this context.

In this paper, we generalize the nonlinear Murnaghan elastic model to account for the nonlinear viscoelastic effects. Our work is motivated by recent experimental studies that have shown a significant increase (by an order of magnitude) in the Murnaghan elastic moduli of some glassy polymers as the frequency of the wave used to measure them decreases~\cite{TechPhys2021}.
This can have a strong effect on the strain wave dynamics, eventually leading to the generation of long soliton-like waves, the spectrum of which is mainly in the low-frequency region. Another important application of this work can be in material assessment since nonlinear moduli can vary significantly even for the same materials but manufactured using different technologies~\cite{TechPhys2021, PoTe2021}.

Many rheological models of nonlinear viscoelastic material have been developed and all of them can be divided into two groups: single- and multiple-integral representations. Single-integral approaches are relatively simple and are widely used for the description of large deformations in polymers~\cite{LaiBakker1995, Schapery1997} and soft biological tissues~\cite{FungBook}, and in nonlinear wave propagation problems~\cite{DePascalis2019, Favrie2023}.
However, the most general framework is provided by the multiple-integral (Green-Rivlin) approach in which stress is expanded in the Fr\'echet series around zero strain history~\cite{GreenRivlin1957, FindleyBook}. This approach has recently attracted researchers' attention for the construction of rheological models~\cite{Lennon2020P1, Lennon2021, Curtis2021}. For the comprehensive overview of different nonlinear viscoelastic models, the reader is referred to reviews~\cite{Drapaca2007, Wineman2009} and chapter~\cite{WardBook_NonlinVisc}.

Based on the Murnaghan elastic model, we intend to build a general viscoelastic model for small but finite strains. The desired generality can only be achieved using the multiple-integral approach.
To derive the expressions for the unknown time- and frequency-dependent nonlinear moduli that arise in this approach, we extend the linear generalized Maxwell constitutive equations written in the internal strains representation~\cite{Banks2008} by including all possible second-order terms.

This paper is organized as follows. We begin with the description of the general finite strain theory in Sec.~\ref{sec:fin_strain_theory}. Here, the equations of motion (Sec.~\ref{sec:eq_mot}) and the Green-Rivlin multiple-integral approach (Sec.~\ref{sec:green_rivlin}) are introduced and the latter is applied to construct the general constitutive equation for the isotropic material. This is followed by the derivation of the general form of time- and frequency-dependent nonlinear dynamic moduli using the internal strains representation in Sec.~\ref{sec:internal_strains}. This completes the derivation of Murnaghan nonlinear viscoelastic model, and in the next section, this model is applied to two important problems related to strain waves in nonlinear solids. Sec.~\ref{sec:prestrain} is devoted to the propagation of small-amplitude waves in prestrained material, which illustrates the model's ability to capture frequency-dependent elastic properties of real materials. In Sec.~\ref{sec:higher_harmonic}, the generation of higher harmonics due to the nonlinear interaction of two strain waves is studied. Here, the dependence of the amplitude of the higher harmonic on nonlinear dynamic moduli is derived, which provides a method to determine all viscoelastic constants of the model. The paper ends with the discussion and conclusion in Secs.~\ref{sec:discussion} and~\ref{sec:conclusion}, respectively.

\section{Finite strain theory}
\label{sec:fin_strain_theory}

\subsection{Equations of motion}
\label{sec:eq_mot}
The behavior of a body is subject to the linear momentum balance equations, which we write in the Cartesian coordinate system:
\begin{equation}
    \label{eq:motion}
    \rho \ddot{\vec{U}} = \divg \mathbf{P},
\end{equation}
where $\vec{U} = \vec{U}(t, \vec{r}) = \vec{U}(t, x, y, z)$ is the displacement vector, $\vec{r}$ is the coordinate vector, the dot denotes the time derivative and $\divg\mathbf{P}$ denotes the divergence of the stress tensor. Throughout this paper we use the material coordinates, therefore, all spatial derivatives are taken with respect to the undeformed configuration, the material density $\rho$ in Eq.~\eqref{eq:motion} is always constant, and $\mathbf{P}$ denotes the first Piola-Kirchhoff stress tensor.

Equation~\eqref{eq:motion} has to be complemented by the stress-strain constitutive equation.
It is convenient to write this equation for the second Piola-Kirchhoff stress tensor~$\mathbf{S}$ since it is symmetric, like the Cauchy stress, due to the balance of angular momentum and is invariant under the rigid body motion. The two stress tensors are related via the deformation gradient as follows:
\begin{equation}
    \mathbf{P} = \bigl(\mathbf{I} + \grad\vec{U}\bigr) \cdot \mathbf{S}.
    \label{eq:P}
\end{equation}
Here, we have already expressed the deformation gradient as the sum of the unit tensor $\mathbf{I}$ and the gradient of displacement, which in index notation writes $\smash{(\grad \vec{U})_{ij} = {\partial U_i}/{\partial r_j}}$, and the central dot denotes the inner (dot) product of two tensors.

\subsection{Green-Rivlin multiple-integral approach}
\label{sec:green_rivlin}
Viscoelastic materials are the materials with memory, i.e. the state of the material at some time is defined by its whole history prior to this time.
In this work we disregard all influences related to external heating and assume that all mechanical interactions in the body can be described by the stress alone and state of the body is completely determined by its strain history.
We suppose that the material is homogeneous and no aging process is taking place so that the stress does not explicitly depend on either the coordinate or the time. Thus, we assume a constitutive relation of the form
\begin{equation}
	\mathbf{S}(t) = \mathbf{S} \left[\mathbfcal{E}(t_1) \big|^{t}_{t_1=-\infty}\right],
	\label{eq:S}
\end{equation}
where the stress tensor $\mathbf{S}$ at each time $t$ is a functional of the strain history prior to~$t$ denoted as $\mathbfcal{E}(t_1) |^{t}_{t_1=-\infty}$ and $\mathbfcal{E}$ is the Green-Lagrange finite strain tensor:
\begin{equation}\label{eq:strain}
	\mathbfcal{E} = \frac12 \Bigl[\grad \vec{U} + \bigl(\grad \vec{U}\bigr)^T + \bigl(\grad \vec{U}\bigr)^T \cdot \grad \vec{U}\Bigr].
\end{equation}

We assume that the material has fading memory, thus, the stress functional in Eq.~\eqref{eq:S} can be expanded as the multiple-integral (Fr\'echet) series~\cite{Pipkin1964}:
\begin{align}
	&\mathbf{S}(t) = \; \mathbf{S}^{(1)}(t) + \mathbf{S}^{(2)}(t) + \dots \nonumber\\[-1.5mm]
	& = \int\limits_{-\infty}^t \mathbf{C}(t-t_1) : \dot{\mathbfcal{E}}(t_1) \mathrm{d}t_1 \nonumber\\[2mm]
	&\hspace{2mm} + \smash{\int\limits_{-\infty}^t\int\limits_{-\infty}^t} \mathbf{N}(t - t_1, t - t_2) :: \dot{\mathbfcal{E}}(t_1) \dot{\mathbfcal{E}}(t_2) \mathrm{d}t_1 \mathrm{d}t_2 + \dots,
	\label{eq:S_expand}
\end{align}
where $\mathbf{C}(t)$ and $\mathbf{N}(t_1, t_2)$ are the fourth- and sixth-order tensorial kernel functions, and dots : and :: between tensors denote double and quadruple contractions, respectively, thus, ${\mathbf{A}:\mathbf{B}}$ denotes $\sum_{ij}A_{\dots ij} B_{ij\dots}$ and $\mathbf{A}::\mathbf{B}\,\mathbf{D}$ denotes $\sum_{ijkl}A_{\dots ijkl} B_{ij\dots} D_{kl\dots}$ in index notation for arbitrary $\mathbf{A}$, $\mathbf{B}$, and $\mathbf{D}$.
We omitted the zeroth order term in Eq.~\eqref{eq:S_expand} since the stress should vanish in the undeformed state, and we assume that initially the material was undeformed: ${\mathbfcal{E}\big|_{t\to-\infty} = 0}$.

In this work we consider small but finite strains, therefore, we keep terms up to the second order in strain and neglect all higher-order terms in Eq.~\eqref{eq:S_expand}.
In what follows, the tensorial kernels $\mathbf{C}(t)$ and $\mathbf{N}(t_1, t_2)$ are referred to as the linear and nonlinear viscoelasticity tensors, respectively. We also note that stress and strain depend on spatial coordinate vector~$\vec{r}$: $\mathbf{S} = \mathbf{S}(t, \vec{r})$ and $\mathbfcal{E} = \mathbfcal{E}(t, \vec{r})$, but we omitted the coordinate in Eqs.~\eqref{eq:S}~--~\eqref{eq:S_expand} for brevity. The viscoelasticity tensors $\mathbf{C}(t)$ and $\mathbf{N}(t_1, t_2)$ are independent of spatial coordinates since we consider homogeneous material.

Arbitrary fourth- and sixth-order tensors have 81 and 729 components, respectively.
The symmetry of the stress and strain tensors ($S_{ij} = S_{ji}$, $\mathcal{E}_{ij} = \mathcal{E}_{ji}$) impose the so-called minor symmetries on the viscoelasticity tensors $C_{ijkl}(t)$ and $N_{ijklmn}(t_1,t_2)$, i.e. they remain unchanged under the swap of indices within each pair $(i,j)$, $(k,l)$, and $(m,n)$.
This reduces the number of independent functions in the linear and nonlinear viscoelasticity tensors to 36 and 216, respectively.
The double integral term in Eq.~\eqref{eq:S_expand} in index notation writes ${\int_{-\infty}^t\int_{-\infty}^t N_{ijklmn}(t - t_1, t - t_2) \dot{\mathcal{E}}_{kl}(t_1) \dot{\mathcal{E}}_{mn}(t_2) \mathrm{d}t_1 \mathrm{d}t_2}$ (Einstein notation for the repeating indices is applied), which implies that the interchange of pairs $(k,l)$ and $(m,n)$ together with the permutation of $t_1$ and $t_2$ does not change the result of this double integral.
Thus, the following symmetry can be assumed:
\begin{equation} \label{eq:N_symm}
	N_{ijklmn}(t_1, t_2) = N_{ijmnkl}(t_2, t_1).
\end{equation}
This reduces the number of independent functions to 126 in the nonlinear viscoelastic tensor. This is still quite a large number, however, it can be further reduced by considering material symmetries, e.g., orthotropic, isotropic, etc.
Sometimes the so-called major symmetries are imposed on the viscoelasticity tensors, i.e. the interchange of pairs $(i,j)$, $(k,l)$, and $(m,n)$ does not affect these tensors. However, this can only be proved for purely elastic materials while for viscoelastic materials, in general, this is not true~\cite{Carcione2014}.

In isotropic material, the viscoelasticity tensors should be invariant under any rotation of the coordinate frame.
This requirement leaves only two independent scalar functions in the linear viscoelasticity tensor.
Isotropy, together with the symmetry condition in Eq.~\eqref{eq:N_symm}, requires that the nonlinear viscoelastic tensor depends on only four scalar functions.
Thus, the expressions for these tensors can be written as follows:
\begin{align}
&C_{ijkl}(t) = K_1(t) \delta_{ij}\delta_{kl} + \frac{K_2(t)}{2} (\delta_{ik}\delta_{jl} + \delta_{il}\delta_{jk}), \label{eq:C_isotrop}\\
&\begin{aligned}[b]
	&N_{ijklmn} (t_1, t_2) = K_3(t_1, t_2) \delta_{ij}\delta_{kl}\delta_{mn}\\
	&+ \frac{K_4(t_1, t_2)}{2} \delta_{ij} (\delta_{km}\delta_{ln} + \delta_{kn}\delta_{lm})
	\\
	&+ \frac{K_5(t_1, t_2)}{2} \delta_{kl} (\delta_{im}\delta_{jn} + \delta_{in}\delta_{jm}) \\
	&+ \frac{K_5(t_2, t_1)}{2} \delta_{mn} (\delta_{ik}\delta_{jl} + \delta_{il}\delta_{jk})\\
	&+ \frac{K_6(t_1, t_2)}{4} \big(
	\delta_{ik}\delta_{jn}\delta_{lm} +
	\delta_{jk}\delta_{in}\delta_{lm} \\
	&\hspace{20mm}+
	\delta_{il}\delta_{jn}\delta_{km} +
	\delta_{ik}\delta_{jm}\delta_{ln} +
	\delta_{jl}\delta_{in}\delta_{km} \\
	&\hspace{20mm}+
	\delta_{jk}\delta_{im}\delta_{ln} +
	\delta_{il}\delta_{jm}\delta_{kn} +
	\delta_{jl}\delta_{im}\delta_{kn}\big),
\end{aligned}
\label{eq:N_isotrop}
\end{align}
where $K_i$ denotes an independent scalar function in the isotropic viscoelastic tensors and $\delta_{ij}$ denotes the Kronecker delta.
Substitution of Eqs.~\eqref{eq:C_isotrop} and~\eqref{eq:N_isotrop} into constitutive equation~\eqref{eq:S_expand} yields the linear and nonlinear parts of stress in isotropic material:
\begin{align}
&\begin{aligned}[b]
	&\mathbf{S}^{(1)}(t) \\&\hspace*{3mm}= \smashoperator{\int\limits_{-\infty}^t} \Big[ K_1(t-t_1) \mathbf{I}\tr\dot{\mathbfcal{E}}(t_1) + K_2(t-t_1)\dot{\mathbfcal{E}}(t_1) \Big] \mathrm{d}t_1,
\end{aligned} \label{eq:S_lin}\\
&\begin{aligned}[b]
	&\mathbf{S}^{(2)}(t) \\&= \smashoperator[l]{\int\limits_{-\infty}^t} \int\limits_{-\infty}^t \Big[
	K_3(t-t_1, t-t_2) \mathbf{I} \tr\dot{\mathbfcal{E}}(t_1) \tr\dot{\mathbfcal{E}}(t_2) \\
	&\hspace{10mm}+ K_4(t-t_1, t-t_2) \mathbf{I} \tr\big(\dot{\mathbfcal{E}}(t_1) \cdot \dot{\mathbfcal{E}}(t_2)\big) \\
	&\hspace{10mm}+ K_5(t-t_1, t-t_2) \dot{\mathbfcal{E}}(t_2) \tr\dot{\mathbfcal{E}}(t_1) \\
	&\hspace{10mm}+ K_5(t-t_2, t-t_1) \dot{\mathbfcal{E}}(t_1) \tr\dot{\mathbfcal{E}}(t_2) \\
	&\hspace{10mm}+ K_6(t-t_1, t-t_2) \Big( \dot{\mathbfcal{E}}(t_1) \cdot \dot{\mathbfcal{E}}(t_2) \\
	&\hspace{36mm}+ \dot{\mathbfcal{E}}(t_2) \cdot \dot{\mathbfcal{E}}(t_1) \Big)
	\Big] \mathrm{d}t_1 \mathrm{d}t_2.
\end{aligned}\label{eq:S_nonlin}
\end{align}
Note, that the third and fourth terms in Eq.~\eqref{eq:S_nonlin}, which involve function $K_5$, make an identical contribution to the stress, therefore, these two terms can be replaced with only one of them multiplied by 2. This operation transforms Eq.~\eqref{eq:S_nonlin} into a form similar to the one given in~\cite{Pipkin1964}.

The symmetry condition~\eqref{eq:N_symm} requires the functions $K_3(t_1,t_2)$, $K_4(t_1,t_2)$, and $K_6(t_1,t_2)$ to be symmetric with respect to the interchange of their arguments. Although symmetry of $K_5(t_1,t_2)$ is sometimes assumed~\cite{FindleyBook}, in general, it is not symmetric.

The choice of six functions for isotropic material is not unique and we are free to replace $K_1$, \dots, $K_6$ with another set of equivalent functions, just as the Lam\'e elastic moduli describing elasticity of an isotropic elastic solid can be replaced with another pair of moduli, e.g., with Young's modulus and Poisson's ratio or bulk and shear moduli. In the following two subsections, we introduce another set of functions instead of $K_1$, \dots, $K_6$ in such a way as to generalize the Murnaghan elastic material to the viscoelastic case.

\subsubsection{Murnaghan elastic material}
\label{sec:murn_elast}
The state of a purely elastic material is completely defined by its specific strain energy~$\Pi$, which for isotropic material undergoing small but finite strains is often written in the form proposed by Murnaghan~\cite{Murnaghan1951, SamsonovBook}:
\begin{align}
	\Pi =\, &\frac{\lambda+2\mu}{2}I_1^2(\mathbfcal{E}) - 2\mu I_2(\mathbfcal{E}) \nonumber \\
	&+ \frac{l + 2m}{3}I_1^3(\mathbfcal{E}) - 2m I_1(\mathbfcal{E}) I_2(\mathbfcal{E}) + n I_3(\mathbfcal{E}),
	\label{eq:pot_en}
\end{align}
Here, $\lambda$ and $\mu$ are the Lam\'e (linear) elastic moduli, $l$, $m$, and $n$ are the Murnaghan (nonlinear) elastic moduli, and $I_1(\mathbfcal{E}) = \tr\mathbfcal{E}$, $I_2(\mathbfcal{E}) = [ \left(\tr\mathbfcal{E}\right)^2 - \tr\mathbfcal{E}^2]/2$, and $I_3(\mathbfcal{E}) = \det \mathbfcal{E}$ denote the invariants of the strain tensor.
This is the most general model of isotropic material since Eq.~\eqref{eq:pot_en} includes all possible second- and third-order terms which are invariant under any rotation of the coordinate frame.
The stress-strain constitutive equation is derived from the strain energy as follows:
\begin{equation}
S^{\text{elast}}_{ij} = \frac{\partial\Pi}{\partial\mathcal{E}_{ij}}.
\label{eq:S_elast_energy}
\end{equation}

The isotropic rheological model defined by Eqs.~\eqref{eq:S_expand}~--~\eqref{eq:S_nonlin} represents the Murnaghan elastic material if
\begin{subequations} \label{eq:K_to_lame_murn_elast}
\begin{align}
	&K_1(t) = \lambda,\\
	&K_2(t) = 2\mu,\\
	&K_3(t_1, t_2) = l - m + \frac{n}{2},\\
	&K_4(t_1, t_2) = m - \frac{n}{2},\\
	&K_5(t_1, t_2) = m - \frac{n}{2},\\
	&K_6(t_1, t_2) = \frac{n}{2}.
\end{align}
\end{subequations}
Here, the functions $K_i$ are constant which reflects the time independence of the elastic properties of a purely elastic material.
In addition, $K_5 = K_4$ due to the requirement that the stress is derived from the scalar potential which is a function of current strain (not a functional of strain history) as shown in Eqs.~\eqref{eq:pot_en} and~\eqref{eq:S_elast_energy}.

We must mention that we use the letters $l$, $m$, and $n$ both as summation indices and as Murnaghan moduli throughout this paper. However, this is the standard notation and it is easy to distinguish between the two cases.

\subsubsection{Murnaghan viscoelastic material}
The elastic properties of a viscoelastic solid are time-dependent, which is reflected by the time-dependent functions $K_1$, \dots $K_6$. In the linear theory, the Lam\'e dynamic moduli $\lambda(t)$ and $\mu(t)$ are usually used instead of the functions $K_1$ and $K_2$. We complement the Lam\'e dynamic moduli with the four Murnaghan dynamic moduli $l(t_1,t_2)$, $m(t_1,t_2)$, $n(t_1,t_2)$, and $h(t_1,t_2)$, which we introduce in the form suggested by Eqs.~\eqref{eq:K_to_lame_murn_elast}:
\begin{subequations} \label{eq:K_to_lame_murn}
	\begin{align}
		&K_1(t) = \lambda(t), \\
		&K_2(t) = 2\mu(t),\\
		&K_3(t_1, t_2) = l(t_1, t_2) - m(t_1, t_2) + \frac{n(t_1, t_2)}{2},\\
		&K_4(t_1, t_2) = m(t_1, t_2) - \frac{n(t_1, t_2)}{2},\\
		&K_5(t_1, t_2) = m(t_1, t_2) - \frac{n(t_1, t_2)}{2} + h(t_1, t_2),\\
		&K_6(t_1, t_2) = \frac{n(t_1, t_2)}{2}.
	\end{align}
\end{subequations}
Note that the need for the fourth modulus $h(t_1, t_2)$ is justified by the fact that Eq.~\eqref{eq:S_elast_energy} is no longer valid and therefore $K_5(t_1, t_2) \neq K_4(t_1, t_2)$.

Since the functions $K_3(t_1, t_2)$, $K_4(t_1, t_2)$, and $K_6(t_1, t_2)$ are symmetric with respect to the interchange of their arguments, as discussed earlier, the dynamic moduli $l$, $m$, and $n$ also possess this property: $l(t_1, t_2) = l(t_2, t_1)$, $m(t_1, t_2) = m(t_2, t_1)$, and $n(t_1, t_2) = n(t_2, t_1)$. However, the dynamic modulus $h$, like $K_5$, in general, is not symmetric: $h(t_1, t_2) \neq h(t_2, t_1)$.

Note that we distinguish between Murnaghan \emph{elastic} moduli, denoted by the constants $l$, $m$, and $n$ in Eq.~\eqref{eq:K_to_lame_murn_elast}, and Murnaghan \emph{dynamic} moduli, denoted by the functions $l(t_1, t_2)$, $m(t_1, t_2)$, $n(t_1, t_2)$, and $h(t_1, t_2)$ in Eq.~\eqref{eq:K_to_lame_murn}. Similarly, we distinguish between Lam\'e elastic moduli, $\lambda$ and $\mu$, and Lam\'e dynamic moduli, $\lambda(t)$ and $\mu(t)$.

Viscoelastic properties of a material are defined by its dynamic moduli, however, there is no commonly accepted functional form for them and the space of suitable functions is infinite-dimensional. In order to measure the dynamic moduli experimentally, one has to choose a finite-dimensional space of functions (e.g. polynomials of a certain degree~\cite{FindleyBook}) and fit a function from this space to experimental results.
In the next subsection, with the help of internal strains formalism, we derive the form of dynamic moduli which is determined by a finite set of constants.

\subsection{Rheological model with internal strains}
\label{sec:internal_strains}

As we have already mentioned, the current state of a viscoelastic material is determined by its strain history. In thermodynamics of continua, deformation processes with memory are often described by a finite number of internal state variables (or memory variables), each of which is subject to an evolution equation. According to this approach, the constitutive equations take the form~\cite{Coleman1967,Maugin1994}:
\begin{align}
\mathbf{S} &= \mathbf{S}(\mathbfcal{E}, \mathbfcal{E}_1, \mathbfcal{E}_2, \dots),\\
\dot{\mathbfcal{E}}_s &= \mathbf{f}_s(\mathbfcal{E}, \mathbfcal{E}_1, \mathbfcal{E}_2, \dots), \quad s = 1, 2,\dots,
\end{align}
where $\mathbfcal{E}_s$ are the internal state variables and the stress function $\mathbf{S}$ together with the functions $\mathbf{f}_s$ define mechanical properties of the material.

One of the most widely used models of a linear viscoelastic solid is the generalized Maxwell model which is also referred to as the Wiechert model or generalized Zener model or generalized standard linear solid~\cite{Banks2008, Favrie2023}.
This model is based on the assumption that multiple simple relaxation processes take place within the material during deformation. Each relaxation process is characterized by its internal strain $\mathbfcal{E}_s$, which plays the role of internal state variable, and relaxation time $\tau_s$. The stress is then assumed to linearly depend on these strains:
\begin{gather}
	\mathbf{S}^{\text{(lin)}}(t) = \sum_{s} \mathbf{C}_{s} : \mathbfcal{E}_{s}(t), \label{eq:S_lin_internal_strains}\\
	\dot{\mathbfcal{E}}_{s} + \frac{\mathbfcal{E}_{s}}{\tau_s} = \dot{\mathbfcal{E}},
	\label{eq:internal_strain_lin}
\end{gather}
where 
$\mathbf{C}_{s}$ denotes constant fourth-order viscoelasticity tensor for corresponding relaxation time.

In this work, we naturally extend the generalized Maxwell model by including all possible next-order corrections to the linear constitutive equations~\eqref{eq:S_lin_internal_strains} and~\eqref{eq:internal_strain_lin}:
\begin{gather}
	\mathbf{S} = \sum_{s} \mathbf{C}_{s} : \tilde{\mathbfcal{E}}_{s} + \sum_{su} \mathbf{N}_{su} :: \tilde{\mathbfcal{E}}_{s} \tilde{\mathbfcal{E}}_{u},
	\label{eq:S_nonlin_internal_strains}\\
	\dot{\tilde{\mathbfcal{E}}}_{s} + \frac{\tilde{\mathbfcal{E}}_{s}}{\tau_s} = \dot{\mathbfcal{E}} + \sum_{uv} \mathbf{B}_{suv} :: \tilde{\mathbfcal{E}}_u \tilde{\mathbfcal{E}}_v.
	\label{eq:internal_strain_nonlin}
\end{gather}
Here, $\tilde{\mathbfcal{E}}_s$ denotes internal strain, which takes into account its nonlinear interaction with other internal strains, the stress depends nonlinearly on internal strains, and the significance of these nonlinear effects is controlled by sixth-order tensors $\mathbf{B}_{suv}$ and $\mathbf{N}_{su}$, respectively.
Note that we can assume the following symmetry:
\begin{align}
(N_{su})_{ijklmn}  &= (N_{us})_{ijmnkl}, \label{eq:Nsu_symm}\\
(B_{suv})_{ijklmn} &= (B_{svu})_{ijmnkl}, \label{eq:Bsuv_symm}
\end{align}
since the terms which these tensors multiply in Eqs.~\eqref{eq:S_nonlin_internal_strains} and~\eqref{eq:internal_strain_nonlin} possess this symmetry.

Equations~\eqref{eq:S_nonlin_internal_strains} and~\eqref{eq:internal_strain_nonlin} can be asymptotically rewritten assuming that the strain and internal strains are small. From this assumption it follows that the difference between the internal strains from Eqs.~\eqref{eq:internal_strain_nonlin} and~\eqref{eq:internal_strain_lin} has the next order of smallness: ${\Delta\mathbfcal{E}_{s} = \tilde{\mathbfcal{E}}_{s} - \mathbfcal{E}_{s} \ll\mathbfcal{E}_{s}}$, which allows us to neglect terms that are nonlinear in $\Delta\mathbfcal{E}_{s}$ or are the product of $\Delta\mathbfcal{E}_{s}$ and $\mathbfcal{E}_{s}$:
\begin{gather}
	\mathbf{S} = \sum_{s} \mathbf{C}_{s} : \big(\mathbfcal{E}_{s} + \Delta\mathbfcal{E}_{s}\big) + \sum_{su} \mathbf{N}_{su} :: \mathbfcal{E}_{s} \mathbfcal{E}_{u},
	\label{eq:S_nonlin_internal_strains2}\\
	\frac{\partial}{\partial t}\Delta\mathbfcal{E}_{s} + \frac{\Delta\mathbfcal{E}_{s}}{\tau_s} =  \sum_{uv} \mathbf{B}_{suv} :: \mathbfcal{E}_{u} \mathbfcal{E}_{v},
	\label{eq:internal_strain_delta}
\end{gather}
Note, that $\mathbfcal{E}_{s}$ evolves according to Eq.~\eqref{eq:internal_strain_lin}, thus, the full set of constitutive equations consists of Eqs.~\eqref{eq:S_nonlin_internal_strains2}, \eqref{eq:internal_strain_delta}, and~\eqref{eq:internal_strain_lin}.

In isotropic material, the fourth-order tensor $\mathbf{C}_s$ and the sixth-order tensors $\mathbf{N}_{su}$ and $\mathbf{B}_{suv}$ (for each $s$, $u$, and $v$) have the same tensorial structure as shown in Eqs.~\eqref{eq:C_isotrop} and~\eqref{eq:N_isotrop}, but with time-independent constants written instead of the functions $K_1$, \dots, $K_6$.

Let us remark on the quasi-static elastic response. This response can be taken into account by requiring that one of the relaxation times be much longer than the characteristic time of the process for which this model is applied. This can be done by making, e.g., $\tau_0$ infinitely large: $\tau_0 \to \infty$. With this assumption, there is no relaxation process for the corresponding internal strains and therefore the following equations hold:
\begin{equation}
\mathbfcal{E}_{0} = \mathbfcal{E}, \quad \Delta\mathbfcal{E}_{0} = 0, \quad \mathbf{B}_{0uv} = 0.
\label{eq:tau0}
\end{equation}
In what follows, we do not separate the terms with infinite $\tau_0$ from other terms with finite relaxation times to make the equations shorter. However, we always imply $\tau_0 \to \infty$ and consequently the relations in Eq.~\eqref{eq:tau0}.

\subsubsection{Multiple-integral form of the rheological model}

In this subsection, we relate the derived nonlinear viscoelastic model to the Green-Rivlin multiple-integral approach described earlier. This allows us to get the general form of the linear and nonlinear viscoelastic tensors which appear in the constitutive equation~\eqref{eq:S_expand}. To do this, we express the internal strains $\mathbfcal{E}_s$ and $\Delta\mathbfcal{E}_s$ in terms of the full strain $\mathbfcal{E}$:
\begin{align}
&\mathbfcal{E}_{s}(t) = \int\limits_{-\infty}^{t} \dot{\mathbfcal{E}}(t_1) e^{-\frac{t-t_1}{\tau_s}} \mathrm{d}t_1.
\label{eq:internal_strain_lin_sln}\\
&\begin{aligned}[b]
	\Delta\mathbfcal{E}_{s}(t) =& \sum_{uv} \int\limits_{-\infty}^{t}\int\limits_{-\infty}^{t} \Big[
	\tilde{\mathbf{B}}_{suv} e^{-\frac{t-t_1}{\tau_u}} e^{-\frac{t-t_2}{\tau_v}} \\
	&- \tilde{\mathbf{B}}_{svu} e^{-\frac{t-t_1}{\tau_s}} e^{-\frac{t_1-t_2}{\tau_u}} \Theta(t_1 - t_2) \\
	&- \tilde{\mathbf{B}}_{suv} e^{-\frac{t-t_2}{\tau_s}} e^{-\frac{t_2-t_1}{\tau_u}} \Theta(t_2 - t_1) \Big] \\
	&\hspace{26mm} :: \dot{\mathbfcal{E}}(t_1) \dot{\mathbfcal{E}}(t_2) \mathrm{d}t_1 \mathrm{d}t_2,
	\label{eq:internal_strain_delta_sln}
\end{aligned}
\end{align}
where $\tilde{\mathbf B}_{suv} = \mathbf{B}_{suv} / \left(\tau_s^{-1} - \tau_u^{-1} - \tau_v^{-1}\right)$. Equation~\eqref{eq:internal_strain_lin_sln} is the straightforward solution of Eq.~\eqref{eq:internal_strain_lin}, while the derivation of Eq.~\eqref{eq:internal_strain_delta_sln} can be found in~\ref{sec:appendix_int_strain_nonlin}.

From the substitution of solutions~\eqref{eq:internal_strain_lin_sln} and~\eqref{eq:internal_strain_delta_sln} into Eq.~\eqref{eq:S_nonlin_internal_strains2}, one can conclude that the linear and nonlinear viscoelastic tensors take the form:
\begin{align}
	\mathbf{C}(t) &= \sum_{s} \mathbf{C}_{s} e^{-\frac{t}{\tau_s}},
	\label{eq:C}\\
	\mathbf{N}(t_1, t_2) &= \sum_{su} \Bigl(  \mathbf{N}^{(1)}_{su} e^{-\frac{t_1}{\tau_s}} e^{-\frac{t_2}{\tau_u}} - \mathbf{N}^{(2)}_{su} E_{su}(t_1,t_2) \nonumber\\
	&\hspace*{27.8mm}- \mathbf{N}^{(3)}_{su} E_{su}(t_2,t_1) \Bigr),
	\label{eq:N}
\end{align}
where the following notation is used:
\begin{align}
	E_{su}(t_1,t_2) &= e^{-\frac{t_1}{\tau_s}} e^{-\frac{t_2-t_1}{\tau_u}} \Theta(t_2 - t_1),
	\label{eq:E}\\
	\mathbf{N}^{(1)}_{su} &= \mathbf{N}_{su} + \sum_{v} \mathbf{C}_v : \tilde{\mathbf{B}}_{vsu},
	\label{eq:N1} \\
	\mathbf{N}^{(2)}_{su} &= \mathbf{C}_s : \sum_{v} \tilde{\mathbf{B}}_{svu},
	\label{eq:N2}\\
	\mathbf{N}^{(3)}_{su} &= \mathbf{C}_s : \sum_{v} \tilde{\mathbf{B}}_{suv},
	\label{eq:N3}
\end{align}
and $\Theta$ denotes the Heaviside step function. The symmetries of the tensors $\mathbf{N}_{su}$ and $\mathbf{B}_{suv}$, written in Eqs.~\eqref{eq:Nsu_symm} and~\eqref{eq:Bsuv_symm}, yield the same symmetry for the tensor~$\mathbf{N}^{(1)}_{su}$, and it follows from Eqs.~\eqref{eq:N2} and~\eqref{eq:N3} that $\mathbf{N}^{(2)}_{su}$ and $\mathbf{N}^{(3)}_{su}$ are related to each other:
\begin{align}
	\big(N^{(1)}_{su}\big)_{ijklmn} = \big(N^{(1)}_{us}\big)_{ijmnkl},
	\label{eq:N1_symm}\\
	\big({N}^{(2)}_{su}\big)_{ijklmn} = \big({N}^{(3)}_{su}\big)_{ijmnkl}.
	\label{eq:N2_N3_symm}
\end{align}

The expression for the linear viscoelastic tensor in the form of a series of decaying exponentials (Eq.~\eqref{eq:C}) is well-known from the literature on the generalized Maxwell model, while the expression for the nonlinear viscoelastic tensor (Eq.~\eqref{eq:N}), to the best of our knowledge, is obtained for the first time.

Tensors $\mathbf{B}_{suv}$, as can be seen from Eqs.~\eqref{eq:N1} -- \eqref{eq:N3}, are included in the expression for the nonlinear viscoelastic tensor $\mathbf{N}(t_1,t_2)$ only in the summed form. This suggests that the number of model parameters can be reduced and the rheological model given by Eqs.~\eqref{eq:S_nonlin_internal_strains2}, \eqref{eq:internal_strain_delta}, and~\eqref{eq:internal_strain_lin} can be defined by tensors $\mathbf{N}^{(1)}_{su}$, $\mathbf{N}^{(2)}_{su}$, and $\mathbf{N}^{(3)}_{su}$ instead of $\mathbf{N}_{su}$ and $\mathbf{B}_{suv}$.

Let us substitute linear and nonlinear viscoelastic tensors given by Eqs.~\eqref{eq:C} and~\eqref{eq:N} into the constitutive equation~\eqref{eq:S_expand}. With the help of internal strains $\mathbfcal{E}_s$ defined in Eq.~\eqref{eq:internal_strain_lin_sln} the result of this substitution writes
\begin{align}
	\mathbf{S}(t) =& \sum_s \mathbf{C}_s : \mathbfcal{E}_s(t) + \sum_{su} \mathbf{N}^{(1)}_{su} :: \mathbfcal{E}_s(t) \mathbfcal{E}_u(t) \nonumber \\
	&- \sum_{su} \smashoperator{\int\limits_{-\infty}^{t}} \mathbf{N}^{(2)}_{su} :: \dot{\mathbfcal{E}}(t_1) \mathbfcal{E}_u(t_1) e^{-\frac{t-t_1}{\tau_s}} \mathrm{d}t_1 \nonumber \\
	&- \sum_{su} \smashoperator{\int\limits_{-\infty}^{t}} \mathbf{N}^{(3)}_{su} ::  \mathbfcal{E}_u(t_1) \dot{\mathbfcal{E}}(t_1)  e^{-\frac{t-t_1}{\tau_s}} \mathrm{d}t_1.
	\label{eq:S_nonlin_internal_strains3}
\end{align}
Now, it is natural to introduce the new internal state variable
\begin{equation}\label{eq:internal_strain_W}
    \mathbfcal{W}_{su}(t) = \int\limits_{-\infty}^{t} \dot{\mathbfcal{E}}(t_1) \otimes \mathbfcal{E}_u(t_1) e^{-\frac{t-t_1}{\tau_s}} \mathrm{d}t_1,
\end{equation}
where $\otimes$ denotes the tensor (outer) product and thus $\mathbfcal{W}_{su}$ is the fourth-order tensor. Using this new variable and taking into account the symmetry in Eq.~\eqref{eq:N2_N3_symm}, equation~\eqref{eq:S_nonlin_internal_strains3} can be written as
\begin{equation}
    \mathbf{S} = \sum_s \mathbf{C}_s : \mathbfcal{E}_s + \smashoperator{\sum_{su}} \bigl( \mathbf{N}^{(1)}_{su} :: \mathbfcal{E}_s \mathbfcal{E}_u - 2\mathbf{N}^{(2)}_{su} :: \mathbfcal{W}_{su}\bigr),
    \label{eq:S_nonlin_internal_strains_new}
\end{equation}
and the new internal variable evolves according to the following equation:
\begin{align}
    \dot{\mathbfcal{W}}_{su} + \frac{\mathbfcal{W}_{su}}{\tau_s} = \dot{\mathbfcal{E}} \otimes \mathbfcal{E}_u.
    \label{eq:internal_strain_nonlin_new}
\end{align}
Once again we mention that $\mathbfcal{E}_s$ is subject to Eq.~\eqref{eq:internal_strain_lin}, thus, the full set of constitutive equations consists of Eqs.~\eqref{eq:S_nonlin_internal_strains_new}, \eqref{eq:internal_strain_nonlin_new}, and~\eqref{eq:internal_strain_lin}.

The reason to use the new above-written form of the model instead of the one given by Eqs.~\eqref{eq:S_nonlin_internal_strains2}, \eqref{eq:internal_strain_delta}, and~\eqref{eq:internal_strain_lin} is that tensors $\mathbf{N}^{(\alpha)}_{su}$, $\alpha=1,2,3$ can be experimentally determined as shown in the following sections for the isotropic material, while tensors $\mathbf{N}_{su}$ and $\mathbf{B}_{suv}$ cannot be unambiguously identified from $\mathbf{N}^{(\alpha)}_{su}$.

Finally, we note the internal strains representation of the rheological model is more suitable for numerical simulation than the integral representation. This is because the direct integration in Eq.~\eqref{eq:S_expand} is computationally too expensive compared to the integration of differential equations~\eqref{eq:internal_strain_lin} and~\eqref{eq:internal_strain_nonlin_new}.

\subsubsection{Isotropic time-dependent moduli}
\label{sec:isotrop_moduli_time}
In isotropic material, the linear viscoelastic tensor depends on two Lam\'e dynamic moduli, which, as follows from Eq.~\eqref{eq:C}, take the form
\begin{subequations}\label{eq:lame_dyn}
	\begin{align}
		&\lambda(t) = \sum\limits_{s} \lambda_s e^{-\frac{t}{\tau_s}},\\
		&\mu(t) = \sum\limits_s \mu_s e^{-\frac{t}{\tau_s}},
	\end{align}
\end{subequations}
where $\lambda_s$ and $\mu_s$ are the viscoelastic constants for corresponding relaxation times $\tau_s$.

The elements of isotropic nonlinear viscoelasticity tensor can be expressed in terms of four dynamic Murnaghan moduli, the form of which can be derived from Eq.~\eqref{eq:N}:
\begin{subequations} \label{eq:murn_dyn}
\begin{align}
&\begin{aligned}[b]
	l(t_1, t_2) = &\sum_{su} \Big[ l^{(1)}_{su} e^{-\frac{t_1}{\tau_s}} e^{-\frac{t_2}{\tau_u}} \\
	&+ l^{(2)}_{su} \left(E_{su}(t_1,t_2) + E_{su}(t_2,t_1)\right) \Big],
\end{aligned} \\
&\begin{aligned}[b]
	m(t_1, t_2) = &\sum_{su} \Big[ m^{(1)}_{su} e^{-\frac{t_1}{\tau_s}} e^{-\frac{t_2}{\tau_u}} \\
	&+ m^{(2)}_{su} \left(E_{su}(t_1,t_2) + E_{su}(t_2,t_1)\right)\Big],
\end{aligned} \\
&\begin{aligned}[b]
	n(t_1, t_2) = &\sum_{su} \Big[ n^{(1)}_{su} e^{-\frac{t_1}{\tau_s}} e^{-\frac{t_2}{\tau_u}} \\
	&+ n^{(2)}_{su} \left(E_{su}(t_1,t_2) + E_{su}(t_2,t_1)\right)\Big],
\end{aligned} \\
&\begin{aligned}[b]
	h(t_1, t_2) = &\sum_{su} \Big[ h^{(1)}_{su} e^{-\frac{t_1}{\tau_s}} e^{-\frac{t_2}{\tau_u}}\\
	&+ h^{(2)}_{su} E_{su}(t_1,t_2) + \tilde h^{(2)}_{su} E_{su}(t_2,t_1)\Big],
\end{aligned}
\end{align}
\end{subequations}
where
$l^{(1)}_{su}$, $m^{(1)}_{su}$, $n^{(1)}_{su}$, and $h^{(1)}_{su}$
denote the isotropic moduli of the sixth-order tensor $\mathbf{N}^{(1)}_{su}$ for each $s$ and $u$, while
$l^{(2)}_{su}$, $m^{(2)}_{su}$, $n^{(2)}_{su}$, $h^{(2)}_{su}$, and $\tilde h^{(2)}_{su}$
denote the isotropic moduli of $\mathbf{N}^{(2)}_{su}$.
For more detailed comments the reader is referred to~\ref{sec:appendix_N}.

The isotropic dynamic moduli, which in general belong to the infinite-dimensional spaces, are now defined in Eqs.~\eqref{eq:lame_dyn} and~\eqref{eq:murn_dyn} by the finite sets of constants, which we refer to as \emph{viscoelastic} moduli: $\lambda_s$ and $\mu_s$ for the Lam\'e dynamic moduli and $l^{(1)}_{su}$, $l^{(2)}_{su}$, $m^{(1)}_{su}$, $m^{(2)}_{su}$, $n^{(1)}_{su}$, $n^{(2)}_{su}$, $h^{(1)}_{su}$, $h^{(2)}_{su}$, and $\tilde h^{(2)}_{su}$ for the Murnaghan dynamic moduli.
The symmetry in Eq.~\eqref{eq:N1_symm} requires $l^{(1)}_{su}$, $m^{(1)}_{su}$, and $n^{(1)}_{su}$ to be symmetric with respect to interchange of indices:
\begin{equation}\label{eq:lmn_symm}
	l^{(1)}_{su} = l^{(1)}_{us}, \quad m^{(1)}_{su} = m^{(1)}_{us}, \quad n^{(1)}_{su} = n^{(1)}_{us},
\end{equation}
which leads to the symmetry of dynamic moduli $l(t_1, t_2)$, $m(t_1, t_2)$, and $n(t_1, t_2)$ with respect to the interchange of $t_1$ and $t_2$ as expected. All other viscoelastic moduli, in general, do not possess this symmetry.
More properties of the viscoelastic moduli are written in~\ref{sec:appendix_moduli}.

\subsubsection{Isotropic frequency-dependent moduli}
\label{sec:isotrop_moduli_freq}
The frequency-dependent dynamic moduli describe the stress response to harmonic strain. Speaking more generally, they determine the stress-strain relation in the frequency domain.
These moduli are defined as follows~\cite{HaoGreenhalgh2019}:
\begin{equation} \label{eq:lame_freq_def}
\lambda(\omega) = -\mathrm{i}\omega \int\limits_{0}^{+\infty} \lambda(t) e^{\mathrm{i}\omega t} dt,
\end{equation}
where upright i denotes the imaginary unit, and the same equation holds for $\mu(\omega)$. The real and imaginary parts of these moduli represent the material's elastic and viscous properties, respectively. Another definition of frequency-dependent moduli exists with the opposite sign of $\omega$ in Eq.~\eqref{eq:lame_freq_def}~\cite{Carcione2014}. Our choice can be justified by the expression for traveling waves $e^{\mathrm{i}k x - \mathrm{i}\omega t}$ which we use in the next section, while the mentioned definition with the opposite sign of $\omega$ is consistent with the traveling waves of the form $e^{\mathrm{i}\omega t - \mathrm{i}k x}$.
We do not want to overcomplicate the notation, so we use the same letter for the time- and frequency dependent moduli. We believe that it will be evident to the reader which one we use from the argument of the moduli which is either time~$t$ or frequency~$\omega$.
The definition in Eq.~\eqref{eq:lame_freq_def} yields the following expressions for the frequency-dependent Lam\'e dynamic moduli
\begin{subequations}\label{eq:lame_freq}
	\begin{align}
		&\lambda(\omega) = -\sum_{s} \frac{\mathrm{i}\, \omega \tau_s \lambda_s}{1 - \mathrm{i}\,\omega\tau_s},\\
		&\mu(\omega) = -\sum_{s} \frac{\mathrm{i}\, \omega \tau_s \mu_s}{1 - \mathrm{i}\,\omega\tau_s}.
	\end{align}
\end{subequations}

It is convenient to introduce the frequency-dependent Murnaghan dynamic moduli in a similar way:
\begin{equation}
	l(\omega_1,\omega_2) = - \omega_1 \omega_2 {\int\limits_{0}^{\infty}} {\int\limits_{0}^{\infty}} l(t_1, t_2) e^{\mathrm{i}\omega_1 t_1} e^{\mathrm{i}\omega_2 t_2} \mathrm{d}t_1 \mathrm{d}t_2,
    \label{eq:murn_freq_def}
\end{equation}
and the same equations hold for $m(\omega_1, \omega_2)$, $n(\omega_1, \omega_2)$, and $h(\omega_1, \omega_2)$.
Substitution of the dynamic moduli given in Eqs.~\eqref{eq:murn_dyn} into Eq.~\eqref{eq:murn_freq_def} yields
\begin{subequations} \label{eq:murn_freq}
\begin{align}
&\begin{aligned}[b]
	l(\omega_1,\omega_2) &= \sum_{su} \Bigl[ l^{(1)}_{su} \mathcal{R}^{(1)}_{su}(\omega_1, \omega_2) \\
	&+ l^{(2)}_{su} \big( \mathcal{R}^{(2)}_{su}(\omega_1, \omega_2) + \mathcal{R}^{(2)}_{su}(\omega_2, \omega_1) \big) \Bigr],
\end{aligned}\\
&\begin{aligned}[b]
	m(\omega_1,\omega_2) &= \sum_{su} \Bigl[ m^{(1)}_{su} \mathcal{R}^{(1)}_{su}(\omega_1, \omega_2) \\
	&+ m^{(2)}_{su} \big( \mathcal{R}^{(2)}_{su}(\omega_1, \omega_2) + \mathcal{R}^{(2)}_{su}(\omega_2, \omega_1) \big) \Bigr],
\end{aligned}\\
&\begin{aligned}[b]
	n(\omega_1,\omega_2) &= \sum_{su} \Bigl[ n^{(1)}_{su} \mathcal{R}^{(1)}_{su}(\omega_1, \omega_2) \\
	&+ n^{(2)}_{su} \big( \mathcal{R}^{(2)}_{su}(\omega_2, \omega_2) + \mathcal{R}^{(2)}_{su}(\omega_2, \omega_1) \big) \Bigr],
\end{aligned}\\
&\begin{aligned}[b]
	h(\omega_1,\omega_2) &= \sum_{su} \Bigl[ h^{(1)}_{su} \mathcal{R}^{(1)}_{su}(\omega_1, \omega_2) \\
	&+ h^{(2)}_{su} \mathcal{R}^{(2)}_{su}(\omega_2, \omega_2) + \tilde h^{(2)}_{su} \mathcal{R}^{(2)}_{su}(\omega_2, \omega_1) \Bigr],
\end{aligned}
\end{align}
\end{subequations}
where the following notation is used:
\begin{subequations} \label{eq:R_su}
	\begin{align}
		&\mathcal{R}^{(1)}_{su}(\omega_1, \omega_2) = -\frac{\omega_1\omega_2 \tau_s \tau_u}{(1 - \mathrm{i}\,\omega_1\tau_s)(1 - \mathrm{i}\,\omega_2\tau_u)}, \\
		&\mathcal{R}^{(2)}_{su}(\omega_1, \omega_2) = -\frac{\omega_1\omega_2 \tau_s \tau_u}{(1 - \mathrm{i}(\omega_1+\omega_2) \tau_s)(1 - \mathrm{i}\,\omega_2\tau_u)}.
	\end{align}
\end{subequations}
The frequency-dependent moduli have similar symmetries as the time-dependent ones, namely, moduli $l(\omega_1,\omega_2),$ $m(\omega_1,\omega_2),$ and $n(\omega_1,\omega_2)$ do not change under the swap of frequencies $\omega_1$ and $\omega_2$.

To conclude this section, let us introduce single and double prime notation for the real and imaginary parts of frequency-dependent dynamic moduli, respectively:
\begin{equation}
	\lambda(\omega) = \lambda'(\omega) + \mathrm{i}\lambda''(\omega),
\end{equation}
and similar expression for $\mu(\omega)$. This notation is widely used for the linear dynamic moduli, while in this article we extend it to the nonlinear dynamic moduli as well: $l(\omega_1, \omega_2) = l'(\omega_1, \omega_2) + \mathrm{i}\, l''(\omega_1, \omega_2)$ and similar expressions for $m(\omega_1, \omega_2)$, $n(\omega_1, \omega_2)$, and $h(\omega_1, \omega_2)$.

\section{Application to wave propagation}

\subsection{Acoustoelastic effect}
\label{sec:prestrain}

In this section, we consider the propagation of small plane waves superimposed upon a static triaxial strain. This problem has been already solved for purely elastic material and it provides a method for experimental measurement of Murnaghan elastic moduli~\cite{HughesKelly53,PoTe2021} based on the change in wave velocity as a function of prestrain (acoustoelastic effect). The recent experimental reports show that the Murnaghan moduli in some types of polystyrene are significantly dependent on the frequency of the periodic wave which is used to measure them~\cite{TechPhys2021}. This suggests that nonlinear viscous effects in polystyrene can be strong and consequently the problem has to be solved for the nonlinear viscoelastic material. Recently, the acoustoelastic effect was considered in the framework of a single-integral approach~\cite{Berjamin2022}, while here we apply the multiple-integral model described in the previous section.

Consider a harmonic strain wave propagating along the $x$ axis in a prestrained material, with the displacement vector taking the form
\begin{equation} \label{eq:displace_with_prestrain}
	U_i = \mathcal{E}^{(0)}_{ii} r_i + A_i e^{\mathrm{i} k x - \mathrm{i} \omega t}.
\end{equation}
Here, $\mathcal{E}^{(0)}_{ii}$ denotes prestretch along axis $i$, which can be viewed as the diagonal component of infinitesimal prestrain tensor $\mathbfcal{E}^{(0)}$, $A_i$ denotes the component of the wave amplitude, $A_i \ll \mathcal{E}^{(0)}_{ii}$ since the wave is assumed to be smaller than the prestrain, $k$ is the wave number, and $\omega$ is the wave frequency.

We substitute the displacement~\eqref{eq:displace_with_prestrain} into the equations of motion~\eqref{eq:motion} complemented by the constitutive equations~\eqref{eq:P} -- \eqref{eq:N_isotrop}, \eqref{eq:K_to_lame_murn}, \eqref{eq:lame_dyn} and~\eqref{eq:murn_dyn} for the nonlinear viscoelastic isotropic material. Since all strains are small and the strain wave is smaller than the prestrain, we neglect all nonlinear terms except for those that are quadratic in $\mathcal{E}^{(0)}_{ii}$ and those that are the product of $A_i$ and $\mathcal{E}^{(0)}_{ii}$. Moreover, we neglect the wave attenuation here, thus the described substitution yields
\begin{subequations} \label{eq:dispersion_prestrain}
	\begin{align}
		&\begin{aligned}[b]
			\rho V_x^2 =& \;\lambda'(\omega) + 2\mu'(\omega) \\& + \tr\mathbfcal{E}^{(0)}(2l'(\omega,0) + \lambda_0 + 2h'(0,\omega))\\ &\hspace{5mm}+ 2\mathcal{E}_{xx}^{(0)} (2m'(\omega,0) + \lambda'(\omega) + 2\mu'(\omega) \\&\hspace{33mm}+ \mu_0 + h'(\omega,0)),
		\end{aligned}\\
		&\begin{aligned}[b]
			\rho V_y^2 = \mu'(\omega) + \tr\mathbfcal{E}^{(0)}\left(\lambda_0 + m'(\omega,0) + h'(0,\omega)\right)\\ + 2\mathcal{E}_{xx}^{(0)} \mu_0 + 2\mathcal{E}_{yy}^{(0)}\mu'(\omega) - \mathcal{E}_{zz}^{(0)}\frac{n'(\omega,0)}{2},
		\end{aligned}\\
		&\begin{aligned}[b]
			\rho V_z^2 = \mu'(\omega) + \tr\mathbfcal{E}^{(0)}\left(\lambda_0 + m'(\omega,0) + h'(0,\omega)\right)\\ + 2\mathcal{E}_{xx}^{(0)} \mu_0 + 2\mathcal{E}_{zz}^{(0)}\mu'(\omega) - \mathcal{E}_{yy}^{(0)}\frac{n'(\omega,0)}{2},
		\end{aligned}
	\end{align}
\end{subequations}
where the prime denotes the real part of a complex-valued modulus, 
$\lambda_0 = \lim_{\omega\to0} \lambda(\omega)$ and $\mu_0 = \lim_{\omega\to0} \mu(\omega)$ are the quasi-static Lam\'e moduli, $l'(\omega, 0) = \lim_{\omega_2\to0} l'(\omega, \omega_2)$ and moduli $m'(\omega, 0)$, $n'(\omega, 0)$, $h'(\omega, 0)$, and $h'(0, \omega)$ have the similar meaning, and $V_i$ denotes phase velocity ($\omega/k$) of the wave polarized along axis $i$.


Following the experimental study~\cite{TechPhys2021}, we consider the material which is prestrained by the pressure~$T$ applied along axis $y$. Since $\mathbfcal{E}^{(0)}$ is the infinitesimal prestrain tensor in Eqs.~\eqref{eq:dispersion_prestrain}, we can use Hooke's law to express the prestrain in terms of the applied pressure:
\begin{equation} \label{eq:prestrain_y}
	\mathcal{E}^{(0)}_{yy} = -T/E_0, \quad \mathcal{E}^{(0)}_{xx} = \mathcal{E}^{(0)}_{zz} = \nu_0 T/E_0,
\end{equation}
where $E_0 = \mu_0(3\lambda_0 + 2\mu_0) / (\lambda_0 + \mu_0)$ is the quasi-static Young's modulus and $\nu_0 = \lambda_0 / (2\lambda_0 + 2\mu_0)$ is the quasi-static Poisson's ratio. Substitution of the prestrain given in Eq.~\eqref{eq:prestrain_y} into the dispersion equations~\eqref{eq:dispersion_prestrain} results in the linear dependence of the wave velocities squared on the applied pressure at a given frequency $\omega$:
\begin{subequations}
\begin{align}
	&\rho V_x^2 = \lambda'(\omega) + 2\mu'(\omega) + b_1(\omega) T,\\
	&\rho V_y^2 = \mu'(\omega) + b_2(\omega) T,\\
	&\rho V_z^2 = \mu'(\omega) + b_3(\omega) T.
\end{align}
\end{subequations}
Here, the coefficients near the pressure $T$ are similar to those given in~\cite{TechPhys2021} but with frequency-dependent moduli:
\begin{subequations} \label{eq:slopes}
\begin{align}
&b_1(\omega) = \frac{ -2l_\text{eff} + (\omega)\frac{\lambda_0}{\mu_0} \bigl(2m_\text{eff}(\omega) + \lambda'(\omega) + 2\mu'(\omega) \bigr)}{3\lambda_0 + 2\mu_0},\\
&b_2(\omega) = \frac{- m_\text{eff}(\omega) - \frac{\lambda_0}{4\mu_0} n_\text{eff}(\omega) - \frac{2(\lambda_0+\mu_0)}{\mu_0} \mu'(\omega)}{3\lambda_0 + 2\mu_0},\\
&b_3(\omega) = \frac{- m_\text{eff}(\omega) + \frac{\lambda_0+\mu_0}{2\mu_0} n_\text{eff}(\omega) + \frac{\lambda_0}{\mu_0} \mu'(\omega)}{3\lambda_0 + 2\mu_0},
\end{align}
\end{subequations}
where the effective Murnaghan dynamic moduli take the form
\begin{subequations} \label{eq:murn_eff_freq}
\begin{align}
	&l_{\text{eff}}(\omega) = l'(\omega,0) + \frac{\lambda_0+\mu_0}{\mu_0} h'(0,\omega) - \frac{\lambda_0}{2\mu_0} h'(\omega,0), \\
	&m_{\text{eff}}(\omega) = m'(\omega,0) + h'(0,\omega),\\
	&n_{\text{eff}}(\omega) = n'(\omega,0).
\end{align}
\end{subequations}
Assuming that $h(t_1, t_2)$ is small compared to the other Murnaghan dynamic moduli and that wave attenuation is small, one obtains that $l(\omega, 0) \approx l_\text{eff}(\omega)$, $m(\omega, 0) \approx m_\text{eff}(\omega)$, and $n(\omega, 0) \approx n_\text{eff}(\omega)$. This allows one to partially determine the Murnaghan dynamic moduli from experiments in a prestrained solid, but a complete measurement of the full set of Murnaghan dynamic moduli is impossible with this experimental procedure.

The exact expressions of the effective Murnaghan elastic moduli in the model with $R$ internal strains take the form
\begin{subequations} \label{eq:murn_eff_freq_exact}
\begin{align}
	&l_{\text{eff}}(\omega) = l_0 + \sum_{s=1}^R \frac{\omega^2 \tau_s^2 l^\text{eff}_s}{1 + \omega^2\tau_s^2},
	\label{eq:l_eff_freq}\\
	&m_{\text{eff}}(\omega) = m_0 + \sum_{s=1}^R \frac{\omega^2 \tau_s^2 m^\text{eff}_s}{1 + \omega^2\tau_s^2},\\
	&n_{\text{eff}}(\omega) = n_0 + \sum_{s=1}^R \frac{\omega^2 \tau_s^2 n^\text{eff}_s}{1 + \omega^2\tau_s^2},
\end{align}
\end{subequations}
where $l_0 = l^{(1)}_{00}$, $m_0 = m^{(1)}_{00}$, and $n_0 = n^{(1)}_{00}$ are the quasi-static Murnaghan moduli and the introduced variables $l^\text{eff}_s$, $m^\text{eff}_s$, and $n^\text{eff}_s$ have the following form:
\begin{subequations}
\begin{align}
	&\begin{multlined}[b]
		l^\text{eff}_s = l^{(1)}_{s0} + l^{(2)}_{s0} + l^{(2)}_{0s} + \frac{\lambda_0+\mu_0}{\mu_0} (h^{(1)}_{0s} + \tilde h^{(2)}_{s0})\\ - \frac{\lambda_0}{2\mu_0}(h^{(1)}_{s0} + h^{(2)}_{s0}),
	\end{multlined}\\
	&m^\text{eff}_s = m^{(1)}_{s0} + m^{(2)}_{s0} + m^{(2)}_{0s} + h^{(1)}_{0s} + \tilde h^{(2)}_{s0},\\
	&n^\text{eff}_s = n^{(1)}_{s0} + n^{(2)}_{s0} + n^{(2)}_{0s}.
\end{align}
\end{subequations}

\begin{figure}
	\includegraphics[width=\linewidth]{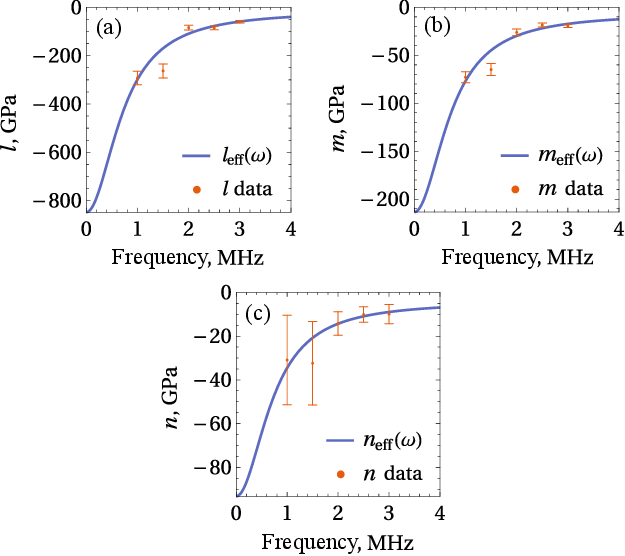}
	\caption{Fitting of the curves in Eqs.~\eqref{eq:murn_eff_freq_exact} with single relaxation time (blue lines) to the experimental data (red points with error bars) for the polystyrene labeled as $\text{PS}_\text{lab2}$ in~\cite{TechPhys2021}. The fitted values are $l_0 = -848$~GPa, $\tilde l_1 = 835$~GPa, $m_0 = -213$~GPa, $\tilde m_1 = 207$~GPa, $n_0 = -93.2$~GPa, $\tilde n_1 = 89.1$~GPa, $\tau_1 = 0.22$~$\mu$s.  \label{fig:murn_fit}}
\end{figure}

We apply the obtained Eqs.~\eqref{eq:murn_eff_freq_exact} to the experimental data shown in Table~2 in~\cite{TechPhys2021} which contains frequency dependence of Murnaghan elastic moduli in different types of polystyrene manufactured using different technologies. In this paper, we use the data for the material labeled $\text{PS}_\text{lab2}$ since it has the most pronounced nonlinear properties.
We use the simplest model with only one relaxation process ($R=1$ in Eqs.~\eqref{eq:murn_eff_freq_exact}) and let $l_0$, $l^\text{eff}_1$, $m_0$, $m^\text{eff}_1$, $n_0$, $n^\text{eff}_1$, and $\tau_1$ be the free parameters so that the effective moduli have their own viscoelastic constants but share the common relaxation time. Experimentally measured values of the moduli at all frequencies are negative (see the data points in Fig.~\ref{fig:murn_fit}) and it seems unlikely that at some frequency they change sign. Thus, we require $l_{\text{eff}}(\omega)$, $m_{\text{eff}}(\omega)$, and $n_{\text{eff}}(\omega)$ to be negative which is insured by $l_0 + l^\text{eff}_1 < 0$ and $l_0 < 0$, and the same constraints for $m$ and $n$. We impose a natural restriction $\tau_1 > 0$ and simultaneously fit the curves in Eqs.~\eqref{eq:murn_eff_freq_exact} using the weighted mean squared error method with inverse squared errors as weights.
The result of this fitting is shown in Fig.~\ref{fig:murn_fit} and the obtained values of free parameters are written in the figure's caption.

The model derived in this paper predicts a further increase in the absolute value of the effective Murnaghan elastic moduli in the considered material as the wave frequency decreases. This suggests that the long waves should exhibit much more significant nonlinear properties than the short ones, which is an interesting result in light of the possible existence of long strain solitons~\cite{WaMot2022}.
We must admit that the fit does not perfectly describe the data since the derived model predicts a smoother change in $l$ and $m$ moduli between 1.5 and 2 MHz than is observed. However,
more experiments in a wider frequency range are needed to assess the accuracy of the derived model when applied to real materials.
The results shown in Fig.~\ref{fig:murn_fit} illustrate the model's ability to describe the material's frequency-dependent nonlinear elastic moduli.

\subsection{Higher harmonic generation}
\label{sec:higher_harmonic}
The experiment in a prestrained body described in the previous subsection has a limited ability to determine the Murnaghan dynamic moduli.
In this section, we study the nonlinear interaction between two harmonic waves, which leads to the creation of a new wave at the sum of frequencies of the two initial waves.
We show that the generated wave provides enough information for a complete measurement of the Murnaghan dynamic moduli.

We apply the standard asymptotic procedure and expand the displacement into power series in a small parameter $\varepsilon$: ${\vec{U} = \varepsilon \vec{U}^{(1)} + \varepsilon^2 \vec{U}^{(2)} + \dots.}$
Substitution of $\vec{U}$ into the equations of motion~\eqref{eq:motion} results in the hierarchy of equations with the linear dissipative wave equation in the leading order and the forced dissipative wave equation in the next order:
\begin{align}
&\rho \ddot{\vec{U}}^{(1)} = \left(\hat\lambda + \hat\mu\right) \grad \divg \vec{U}^{(1)} + \hat\mu \nabla^2 \vec{U}^{(1)},
\label{eq:motion1}\\
&\begin{multlined}[b]
	\rho \ddot{\vec{U}}^{(2)} = \left(\hat\lambda + \hat\mu\right) \grad \divg \vec{U}^{(2)} + \hat\mu \nabla^2 \vec{U}^{(2)} + \vec{F}\big[\vec{U}^{(1)}\big].
\end{multlined}
\label{eq:motion2}
\end{align}
Here, $\nabla^2$ denotes the Laplace operator, $\hat\lambda$ and $\hat\mu$ denote the retarded integral operators which act on an arbitrary function $f(\vec{r},t)$ in the following way:
\begin{subequations}
\begin{align}
	\hat\lambda f(\vec{r},t) = \sum_s \lambda_s \int\limits_{-\infty}^t \dot f(\vec{r}, t') e^{-\frac{t-t'}{\tau_s}} \mathrm{d}t', \\
	\hat\mu f(\vec{r},t) = \sum_s \mu_s \int\limits_{-\infty}^t \dot f(\vec{r}, t') e^{-\frac{t-t'}{\tau_s}} \mathrm{d}t',
\end{align}
\end{subequations}
and $\vec{F}\big[\vec{U}^{(1)}\big]$ is the nonlinear operator of $\vec{U}^{(1)}$ which acts as a force on $\vec{U}^{(2)}$. Its expression takes the form
\begin{multline}
\vec{F}[U^{(1)}] = \divg \Big(\mathbf{S}^{(1)}[\mathbfcal{E}^\text{nl}] + \grad \vec{U}^{(1)} \cdot \mathbf{S}^{(1)}[\mathbfcal{E}^\text{lin}] \\+ \mathbf{S}^{(2)}[\mathbfcal{E}^\text{lin}]\Big),
\end{multline}
where the following notation is used:
\begin{align}
\mathbfcal{E}^\text{lin} &= \frac12 \Bigl[\grad \vec{U}^{(1)} + \bigl(\grad \vec{U}^{(1)}\bigr)^T \Bigr],\\
\mathbfcal{E}^\text{nl} &= \frac12 \bigl(\grad \vec{U}^{(1)}\bigr)^T \cdot \grad \vec{U}^{(1)},
\end{align}
and the expressions for $\mathbf{S}^{(1)}[\mathbfcal{E}^\text{lin}]$, $\mathbf{S}^{(1)}[\mathbfcal{E}^\text{nl}]$, and $\mathbf{S}^{(2)}[\mathbfcal{E}^\text{nl}]$ are given in Eqs.~\eqref{eq:S_lin} and~\eqref{eq:S_nonlin} with $\mathbfcal{E}^\text{lin}$ or $\mathbfcal{E}^\text{nl}$ substituted instead of $\mathbfcal{E}$.

The leading order equation~\eqref{eq:motion1} has the general solution in the form of a decaying harmonic wave.
Let us consider the solution given by the sum of two decaying harmonic waves with either $\vec{k}$ or $\omega$ having a nonzero imaginary part responsible for attenuation:
\begin{equation} \label{eq:U1}
	\vec{U}^{(1)} = \vec{A}_{1} e^{\mathrm{i} \vec{k}_{1} \cdot \vec{r} - \mathrm{i}\omega_1 t} + \vec{A}_{2} e^{\mathrm{i} \vec{k}_{2} \cdot \vec{r} - \mathrm{i}\omega_2 t} + \text{c.c.}
\end{equation}
The nonlinear force $\vec{F}\big[\vec{U}^{(1)}\big]$ in Eq.~\eqref{eq:motion2} consists of decaying harmonic forces with doubled wave vectors and frequencies of each wave and also with their sum and difference. Here, we focus on the force with the sum of wave vectors and frequencies, which has the form:
\begin{equation}
	\vec{F}\big[\vec{U}^{(1)}\big] = \vec{A}_{F} e^{\mathrm{i} \vec{k}_{3} \cdot \vec{r} - \mathrm{i}\omega_3 t}  + \text{c.c.},
\end{equation}
where $\omega_3 = \omega_1 + \omega_2$, $\vec{k}_{3} = \vec{k}_{1} + \vec{k}_{2}$, force amplitude $\vec{A}_F$ is a function of $\vec{A}_1$, $\vec{A}_2$, $\vec{k}_1$, $\vec{k}_2$, $\omega_1$, and $\omega_2$, and has a lengthy expression which we do not write here for brevity. This force will excite the wave with the same frequency and wave vector, but with different amplitude:
\begin{equation} \label{eq:U2}
	\vec{U}^{(2)} = \vec{A}_{3} e^{\mathrm{i} \vec{k}_{3} \cdot \vec{r} - \mathrm{i}\omega_3 t}  + \text{c.c.}
\end{equation}
Equation~\eqref{eq:motion2} provides the resonance relation between the amplitude of the generated wave $\vec{A}_3$ and the amplitude of the force $\vec{A}_{F}$ in the form
\begin{align}
	\vec{A}_3 = \mathbf{M}^{-1} \cdot \vec{A}_{F},
\end{align}
where $\mathbf{M}$ is the 3 by 3 matrix defined as
\begin{equation}
	\mathbf{M} = \left(k_3^2 \, \mu(\omega_3) - \rho\omega_3^2\right) \mathbf{I} + \vec{k}_3 \otimes \vec{k}_3 \left(\lambda(\omega_3) + \mu(\omega_3)\right).
\end{equation}

We apply the obtained results to waves propagating in different directions and with different polarizations (amplitude direction). The four simplest combinations of wave vectors and polarizations of the two waves
are summarized in Table~\ref{tab:waves}.
In all of the listed cases, the resulting wave has single polarization which is either longitudinal or transverse, but not mixed.
In these cases, the amplitude of the resulting wave $\vec{A}_3$ can be expressed as follows:
\begin{equation} \label{eq:a_res}
	\vec{A}_3 = -\frac{\mathrm{i} k_1 k_2 (k_1 + k_2) A_1 A_2}{\left(k_3^2 \, D - \rho\omega_3^2\right)} \vec{a}_F,
\end{equation}
where $\vec{a}_{F}$ and $D$, which depend only on the dynamic moduli, are written in Table~\ref{tab:waves} and the scalar values $k_1$, $k_2$, $k_3$, $A_1$, and $A_2$ denote the magnitudes of the corresponding vectors.
In the first three cases (No.~1 -- 3), both waves propagate in the same direction along axis $x$ with wave vectors $\vec{k}_1 = k_1 \vec{e}_x$ and $\vec{k}_2 = k_2 \vec{e}_x$, where $\vec{e}_x$ is the $x$ unit vector. The last case (No. 4) corresponds to the perpendicularly propagating waves with $\vec{k}_1 = k_1 \vec{e}_x$ and $\vec{k}_2 = k_2 \vec{e}_y$. In each of the four cases shown in Table~\ref{tab:waves}, the initial waves are polarized along a single axis, e.g. in case No.~2: $\vec{A}_1 = A_1 \vec{e}_x$ and $\vec{A}_2 = A_2 \vec{e}_y$.
\begin{table*}
	\centering
	\begin{tabular}{ccccccc}
		No. & $\vec{k}_1$ & $\vec{A}_1$ & $\vec{k}_2$ & $\vec{A}_2$ & $\vec{a}_F$ & $D$ \\
		\hline
		1 & $x$ & $x$ & $x$ & $x$ & $\displaystyle\begin{aligned}
			\vec{e}_x \big[& 2h(\omega_1,\omega_2) + 2h\left(\omega_2,\omega_1\right) + 2l(\omega_1,\omega_2) + 4m(\omega_1,\omega_2)\\
			&+\lambda (\omega_1) + \lambda (\omega_2) + \lambda(\omega_3)
			+ 2\mu(\omega_1) + 2\mu(\omega_2) + 2\mu(\omega_3)\big]
		\end{aligned}$ & $\lambda(\omega_3) + 2\mu(\omega_3)$ \\
		2 & $x$ & $x$ & $x$ & $y$ & $
		\displaystyle\vec{e}_y \left[ h(\omega_1,\omega_2) + m(\omega_1,\omega_2) + \lambda(\omega_2) + 2\mu(\omega_2) \right]$ & $\mu(\omega_3)$ \\
		3 & $x$ & $y$ & $x$ & $y$ & $
		\displaystyle \vec{e}_x \left[m(\omega_1,\omega_2) + \lambda(\omega_3) + 2\mu(\omega_3) \right]$ & $\lambda(\omega_3) + 2\mu(\omega_3)$ \\
		4 & $x$ & $y$ & $y$ & $z$ & $\displaystyle \vec{e}_z \frac{k_1}{k_1 + k_2} \left[\frac{n(\omega_1,\omega_2)}{4} + \mu(\omega_2)\right] $ & $\mu(\omega_3)$ \\
	\end{tabular}
	\caption{\label{tab:waves} Generation of a harmonic at the sum of frequencies of the two waves. The columns $\vec{k}_1$, $\vec{A}_1$, $\vec{k}_2$ and $\vec{A}_2$ indicate the direction of these vectors, e.\,g., $x$ in the $\vec{k}_1$ column indicates that $\vec{k}_1 = k_1 \vec{e}_x$ and $y$ in the $\vec{A}_2$ column indicates that $\vec{A}_2 = A_2 \vec{e}_y$. Equation~\eqref{eq:a_res} should be used to obtain the amplitude of the generated higher harmonic wave. As mentioned in the text, $\omega_3 = \omega_1 + \omega_2$.}
\end{table*}

The results shown in Table~\ref{tab:waves} suggest that all Murnaghan dynamic moduli can be obtained by the generation of two harmonic waves with different wave vectors $\vec{k}_1$ and $\vec{k}_2$ and frequencies $\omega_1$ and $\omega_2$ and measuring the amplitude of the higher harmonic with $\vec{k}_1 + \vec{k}_2$ and $\omega_1 + \omega_2$. The dynamic moduli $l$, $m$, and $h$ can be obtained from experiments No.~1~--~3, where both initial waves and the higher harmonic propagate along the same axis. To obtain the dynamic modulus $n$ one has to use a more sophisticated technique (experiment No.~4) and generate perpendicularly propagating waves and measure the higher-frequency wave traveling in the direction between the two initial waves.

\section{Discussion}
\label{sec:discussion}

The goal of the present study, as indicated in the Introduction, is to generalize the Murnaghan elastic material to account for nonlinear viscoelastic properties. We sought to derive the most general model in the small but finite strain regime which justifies the choice of the Green-Rivlin multiple-integral approach.
This approach is often said to be too general and complex compared to the single-integral approaches for the description of large strains. This is due to the rapid increase in the number of unknown functions (dynamic moduli) when the higher-order terms are included in the constitutive equation~\eqref{eq:S_expand}.
However, in this paper, we tried to show that the complexity of multiple-integral approach is plausible in case of small strains.

In the present article, we considered the constitutive equation~\eqref{eq:S_expand} in the most general tensorial form provided by Fr\'echet series expansion. Sometimes, the scalar multiple-integral expansion or Volterra series can be used, e.g. for the description of shear stress in viscoelastic fluids undergoing simple shear deformation~\cite{Lennon2020P1,Lennon2020P2}. This significantly reduces the number of dynamic moduli, namely to a single modulus in each order. However, stress is essentially a tensor that requires a tensorial constitutive equation.

In general, a viscoelastic material with the constitutive equation~\eqref{eq:S_expand} is characterized by the 36 linear and 126 second-order (nonlinear) dynamic moduli. Linear dynamic moduli depend on a single time or frequency variable, while nonlinear dynamic moduli are defined in two-dimensional time or frequency space. The number of dynamic moduli is significantly reduced when material symmetry is taken into account. In the simplest case of isotropic material, there are only two linear (Lam\'e) and four nonlinear (Murnaghan) dynamic moduli.

The state of a viscoelastic material is defined by its whole strain history. In thermodynamics, it is often assumed that the state of a body during a process with memory is determined by a set of internal state variables, each of which is subject to an evolution equation.
We applied this approach to deduce the general form of the dynamic moduli which arise in the multiple-integral expansion. We started from the linear generalized Maxwell model of elasticity and then added all possible second-order terms in its constitutive equations. Application of small strain assumption and solving the evolution equations for the internal variables (internal strains) allowed us to derive the general form of linear and nonlinear dynamic moduli. The expression for the linear moduli (Eq.~\eqref{eq:C}) coincides with the one obtained in the framework of generalized Maxwell model while the expression for the nonlinear moduli (Eq.~\eqref{eq:N}) is obtained for the first time. With the help of these general expressions, the isotropic (Lam\'e and Murnaghan) time- and frequency-dependent moduli are obtained (Eqs.~\eqref{eq:lame_dyn}, \eqref{eq:murn_dyn}, \eqref{eq:lame_freq}, and~\eqref{eq:murn_freq}).

The derived expressions depend on a finite number of constants. In isotropic material, these include the Lam\'e viscoelastic moduli $\lambda_s$ and $\mu_s$, the Murnaghan viscoelastic moduli $l^{(1)}_{su}$, $l^{(2)}_{su}$, $m^{(1)}_{su}$, $m^{(2)}_{su}$, $n^{(1)}_{su}$, $n^{(2)}_{su}$, $h^{(1)}_{su}$, $h^{(2)}_{su}$, and $\tilde h^{(2)}_{su}$, and the relaxation times $\tau_s$.
The number of Lam\'e viscoelastic moduli depends linearly and the number of Murnaghan viscoelastic moduli depends quadratically on the number of relaxation processes. This can lead to a large number of model parameters, especially if many relaxation processes are included in the model. However, the number of parameters is somewhat reduced if several restrictions are taken into account, such as the symmetry of $l^{(1)}_{su}$, $m^{(1)}_{su}$, and $n^{(1)}_{su}$ with respect to the interchange of indices and other relations described in Appendix~\ref{sec:appendix_moduli}.

We applied the derived rheological model to the problem of small-amplitude wave propagation in a prestrained solid (Sec.~\ref{sec:prestrain}).  The importance of this problem lies in the fact that it provides a method for the experimental measurement of the Murnaghan elastic moduli of a purely elastic material. In the case of a viscoelastic material, we showed that it does not allow one to measure all the Murnaghan viscoelastic moduli. However, the derived expressions of the Murnaghan dynamic moduli explained the frequency dependence of the effective Murnaghan moduli observed in the experiments.
The considered problem also illustrates that not all viscoelastic moduli are required to describe some specific wave processes and simpler models can be obtained from the general approach presented in this article.

Another important nonlinear wave problem is the generation of higher harmonic which we considered in Sec.~\ref{sec:higher_harmonic}. Two harmonic waves with different frequencies $\omega_1$ and $\omega_2$ generate the new wave at $\omega_1 + \omega_2$ frequency, the amplitude of which depends on nonlinear dynamic moduli at ($\omega_1, \omega_2$) and ($\omega_2, \omega_1$) points in the two-dimensional frequency space. This allows one to fully investigate the frequency dependence of the Murnaghan dynamic moduli, and we showed that each modulus can be measured in this way. We must mention that the idea of identifying higher-order dynamic moduli using nonlinear wave interaction was used in recent works~\cite{Lennon2020P1,Lennon2020P2}.


One of the future research directions is the study of soliton-like waves in nonlinear viscoelastic materials. Strain solitons and soliton-like waves continue to attract researchers' attention, and so far these waves have been studied either within the framework of linear viscosity or without viscous effects at all.
Another continuation of this work is the determination of elastic properties of nanostructured materials from the known properties of matrix and nanoinclusions. This problem has been already solved for absolutely elastic materials~\cite{SemenovBeltukov2020}, however, the matrix of nanostructured materials is often made of viscoelastic glassy polymers. Therefore, the extension of the work~\cite{SemenovBeltukov2020} to account for the viscoelastic matrix and determination of nonlinear dynamic moduli of nanostructured material is an important problem for future work.

\section{Conclusion}
\label{sec:conclusion}
The general second-order nonlinear viscoelastic model is derived using the Green-Rivlin multiple-integral approach.
In the isotropic material, this approach yields the four scalar time-dependent functions (Murnaghan dynamic moduli) which describe the material's second-order viscoelastic properties.
The general form of time- and frequency-dependent nonlinear dynamic moduli is obtained using the systematic extension of the generalized Maxwell model by including all possible second-order corrections into its constitutive equations written in internal strains representation. This approach yielded the differential form of the nonlinear viscoelastic model which is preferable to the integral form for numerical simulations.

The derived model is applied to the problem of infinitesimal strain wave propagation in a finitely prestrained material, which allows to measure the Murnaghan elastic moduli due to the acoustoelastic effect. The obtained frequency-dependent expressions of the Murnaghan dynamic moduli describe the pronounced frequency dependence of the effective Murnaghan elastic moduli observed in experiments. It is shown that these experiments are not sufficient to measure all viscoelastic constants in the derived model.
To determine all these constants in an isotropic material, the nonlinear interaction of two harmonic strain waves of different polarizations propagating in different directions is considered. This interaction generates higher frequency harmonics, the amplitudes of which depend on nonlinear dynamic moduli, providing a method for their determination.

\section*{Acknowledgements}
	This work was supported by the Russian Science Foundation, project no.~22-72-10083.


\appendix

\section{Additional internal strains}
\label{sec:appendix_int_strain_nonlin}
The solution to Eq.~\eqref{eq:internal_strain_delta} has the form
\begin{equation}
	\Delta\mathbfcal{E}_{s} = \sum_{uv} \int\limits_{-\infty}^{t} \mathbf{B}_{suv} :: \mathbfcal{E}_{u}(t_1) \mathbfcal{E}_{v}(t_1) e^{-\frac{t-t_1}{\tau_s}} \mathrm{d}t_1.
	\label{eq:int_strain_delta_sln1}
\end{equation}
Substitution of solution~\eqref{eq:internal_strain_lin_sln} into Eq.~\eqref{eq:int_strain_delta_sln1} yields
\begin{align}
	\Delta\mathbfcal{E}_{s} = \sum_{uv} &\int\limits_{-\infty}^t \int\limits_{-\infty}^t \int\limits_{-\infty}^{t} \mathbf{B}_{suv} ::  \dot{\mathbfcal{E}}(t_2) \dot{\mathbfcal{E}}(t_3) \nonumber\\
	&\times e^{-\frac{t-t_1}{\tau_s}} e^{-\frac{t_1-t_2}{\tau_u}} e^{-\frac{t_1-t_3}{\tau_v}} \nonumber\\
	&\times \Theta(t_1 - t_2) \Theta(t_1 - t_3) \mathrm{d}t_1 \mathrm{d}t_2 \mathrm{d}t_3,
	\label{eq:int_strain_delta_sln2}
\end{align}
where the Heaviside theta functions denote that the integration with respect to $t_2$ and $t_3$ is done from negative infinity to $t_1$.
This equation can be integrated with respect to $t_1$ using the following auxiliary calculations:
\begin{align}
	&\int\limits_{-\infty}^t e^{-\frac{t-t_1}{\tau_s}} e^{-\frac{t_1-t_2}{\tau_u}} e^{-\frac{t_1-t_3}{\tau_v}} \Theta(t_1 - t_2) \Theta(t_1 - t_3) \mathrm{d}t_1 \nonumber\\
	&= \smashoperator[r]{\int\limits_{\max(t_2,t_3)}^t} e^{-\frac{t-t_1}{\tau_s}} e^{-\frac{t_1-t_2}{\tau_u}} e^{-\frac{t_1-t_3}{\tau_v}} \mathrm{d}t_1 \nonumber\\
	&= \frac{1}{\frac1{\tau_s} - \frac1{\tau_u} - \frac1{\tau_v}} \Big(e^{-\frac{t-t_2}{\tau_u}} e^{-\frac{t-t_3}{\tau_v}} \nonumber\\
	&\hspace{5mm}- e^{-\frac{t-\max(t_2,t_3)}{\tau_s}} e^{-\frac{\max(t_2,t_3)-t_2}{\tau_u}} e^{-\frac{\max(t_2,t_3)-t_3}{\tau_v}}\Big).
\end{align}
The second term in the brackets can be rewritten as
\begin{align}
	&e^{-\frac{t-\max(t_2,t_3)}{\tau_s}} e^{-\frac{\max(t_2,t_3)-t_2}{\tau_u}} e^{-\frac{\max(t_2,t_3)-t_3}{\tau_v}} \nonumber\\
	&= e^{-\frac{t-t_2}{\tau_s}} e^{-\frac{t_2-t_3}{\tau_v}} \Theta(t_2 - t_3) + e^{-\frac{t-t_3}{\tau_s}} e^{-\frac{t_3-t_2}{\tau_u}} \Theta(t_3 - t_2).
\end{align}
With the derivations shown above Eq.~\eqref{eq:int_strain_delta_sln2} takes the form
\begin{align}
	\Delta\mathbfcal{E}_{s} = &\sum_{uv} \int\limits_{-\infty}^{t}\int\limits_{-\infty}^{t} \bigg[
	\tilde{\mathbf{B}}_{suv} e^{-\frac{t-t_2}{\tau_u}} e^{-\frac{t-t_3}{\tau_v}} \nonumber\\
	&- \tilde{\mathbf{B}}_{suv} e^{-\frac{t-t_2}{\tau_s}} e^{-\frac{t_2-t_3}{\tau_v}} \Theta(t_2 - t_3) \nonumber\\
	&- \tilde{\mathbf{B}}_{suv} e^{-\frac{t-t_3}{\tau_s}} e^{-\frac{t_3-t_2}{\tau_u}} \Theta(t_3 - t_2) \bigg] \nonumber\\
	&\hspace{27mm} :: \dot{\mathbfcal{E}}(t_2) \dot{\mathbfcal{E}}(t_3) \mathrm{d}t_2 \mathrm{d}t_3,
\end{align}
where $\tilde{\mathbf{B}}_{suv} = \mathbf{B}_{suv} / \left(\frac1{\tau_s} - \frac1{\tau_u} - \frac1{\tau_v}\right)$ as mentioned in the main text. Finally, we obtain the~Eq.~\eqref{eq:internal_strain_delta_sln} by the interchange of the indices $u$ and $v$ in the second term in the brackets and renaming $t_2$ and $t_3$ to $t_1$ and $t_2$, respectively.

\section{Isotropic tensors}
\label{sec:appendix_N}
The isotropic sixth-order tensors $\mathbf{N}^{(\alpha)}_{su}$ from Eqs.~\eqref{eq:N1} and~\eqref{eq:N2} take the form
\begin{align}
\big(&N^{(\alpha)}_{su}\big)_{ijklmn} = \left(l_{su}^{(\alpha)} - m_{su}^{(\alpha)} + \frac{n_{su}^{(\alpha)}}{2}\right) \delta_{ij}\delta_{kl}\delta_{mn} \nonumber \\
&+ \frac12 \left(m_{su}^{(\alpha)} - \frac{n_{su}^{(\alpha)}}{2}\right) \delta_{ij} (\delta_{km}\delta_{ln} + \delta_{kn}\delta_{lm})
 \nonumber \\
&+ \frac12 \left(m_{su}^{(\alpha)} - \frac{n_{su}^{(\alpha)}}{2} + h^{(\alpha)}_{su}\right) \delta_{kl} (\delta_{im}\delta_{jn} + \delta_{in}\delta_{jm})  \nonumber \\
&+ \frac12 \left(m_{su}^{(\alpha)} - \frac{n_{su}^{(\alpha)}}{2} + \tilde h^{(\alpha)}_{su}\right) \delta_{mn} (\delta_{ik}\delta_{jl} + \delta_{il}\delta_{jk}) \nonumber \\
&+ \frac{n_{su}^{(\alpha)}}{8} \Big(
\delta_{ik}\delta_{jn}\delta_{lm} +
\delta_{jk}\delta_{in}\delta_{lm}  \nonumber \\
&\hspace{13mm}+
\delta_{il}\delta_{jn}\delta_{km} +
\delta_{ik}\delta_{jm}\delta_{ln} +
\delta_{jl}\delta_{in}\delta_{km}  \nonumber \\
&\hspace{13mm}+
\delta_{jk}\delta_{im}\delta_{ln} +
\delta_{il}\delta_{jm}\delta_{kn} +
\delta_{jl}\delta_{im}\delta_{kn}\Big),
\label{eq:N_alpha}
\end{align}
where $l_{su}^{(\alpha)},$ $m_{su}^{(\alpha)},$ $n_{su}^{(\alpha)},$ $h_{su}^{(\alpha)},$ and $\tilde h_{su}^{(\alpha)}$ are the five arbitrary constants (viscoelastic moduli) for each $s$ and $u$. The symmetry in Eq.~\eqref{eq:N1_symm} requires the viscoelastic moduli $l_{su}^{(1)},$ $m_{su}^{(1)}$ and $n_{su}^{(1)}$ to be symmetric with respect to interchange of indices, as mentioned in Eq.~\eqref{eq:lmn_symm}, and $\tilde h_{su}^{(1)} = h_{us}^{(1)}$. Thus, only four independent sets of viscoelastic moduli define the set of isotropic tensors $\mathbf{N}^{(1)}_{su}$. Tensors $\mathbf{N}^{(2)}_{su}$ do not possess any additional symmetries, thus, all the five sets of viscoelastic moduli $l_{su}^{(2)},$ $m_{su}^{(2)},$ $n_{su}^{(2)},$ $h_{su}^{(2)}$, and $\tilde h_{su}^{(2)}$ are required to describe $\mathbf{N}^{(2)}_{su}$ in the isotropic case.

Substitution of Eq.~\eqref{eq:N_alpha} into Eq.~\eqref{eq:N} with the discussed symmetries for $\alpha=1$ leads to the expression defined by Eqs.~\eqref{eq:N_isotrop} and~\eqref{eq:K_to_lame_murn}, where the Murnaghan dynamic moduli take the form given in Eqs.~\eqref{eq:murn_dyn}.

\section{Properties of dynamic moduli}
\label{sec:appendix_moduli}
The dynamic moduli in general are arbitrary functions that are defined for all non-negative times and determine the material mechanical properties. However, moduli should satisfy certain conditions which ensure that the material behaves properly. In this subsection, we mention some of these properties.

First, consider the material which undergoes a step strain at time $t=0$ and another step strain at time $t=\tau$ ($\tau$ is arbitrary), and then its strain remains unchanged. This material should relax to its static elastic state, thus the following limits hold:
\begin{align}
	&\lim\limits_{t\to\infty}\lambda(t) = \lambda_0, & &\lim\limits_{t\to\infty}\mu(t) = \mu_0, \\
	&\lim\limits_{t\to\infty}l(t,t-\tau) = l_0, & &\lim\limits_{t\to\infty}m(t,t-\tau) = m_0, \\
	&\lim\limits_{t\to\infty}n(t,t-\tau) = n_0, & &\lim\limits_{t\to\infty}h(t,t-\tau) = 0.
\end{align}
The non-linear quasi-static moduli $l_0$, $m_0$, and $n_0$ can take arbitrary values either positive or negative, while the fourth modulus $h$ has to be zero in the quasi-static elastic limit since isotropic elastic material has only three second-order nonlinear moduli as discussed in Sec.~\ref{sec:murn_elast}. Similar limits should hold for the frequency-dependent moduli at $\omega\to 0$.

Second, the instantaneous response of a material on a sudden deformation must be purely elastic:
\begin{align}
	&\lim_{t\to0}\lambda(t) = \lambda_\infty, & & \lim_{t\to0}\mu(t) = \mu_\infty,\\
	&\lim_{\substack{t_1\to0\\t_2\to0}} l(t_1,t_2) = l_\infty, & & \lim_{\substack{t_1\to0\\t_2\to0}} m(t_1,t_2) = m_\infty,\\
	&\lim_{\substack{t_1\to0\\t_2\to0}} n(t_1,t_2) = n_\infty, & & \lim_{\substack{t_1\to0\\t_2\to0}} h(t_1,t_2) = 0.
\end{align}
Here, in the instantaneous elastic limit, the modulus $h$ must vanish. Similar limits should hold for the frequency-dependent moduli at $\omega_1\to\infty$ and $\omega_2\to\infty$.
We denote static moduli with subscript 0 and instantaneous moduli with subscript $\infty$ because they correspond to zero and infinite frequency processes, respectively.

The last property, which we want to mention, is that the instantaneous response of a prestrained material (strained material which reached its static state) should be elastic as well:
\begin{align}
	\lim_{\substack{t_1\to0\\t_2\to\infty}} h(t_1,t_2) = \lim_{\substack{t_1\to\infty\\t_2\to0}} h(t_1,t_2) = 0.
\end{align}

The above-written equations impose certain restrictions on the moduli matrices, which define time- and frequency-dependent moduli in Eqs.~\eqref{eq:murn_dyn} and~\eqref{eq:murn_freq}, respectively:
\begin{align}
	&l^{(1)}_{00} = l_0, \quad m^{(1)}_{00} = m_0, \quad n^{(1)}_{00} = n_0, \quad h^{(1)}_{00} = 0, \\
	&\sum_{su} l^{(1)}_{su} + l^{(2)}_{su} = l_\infty,\\
	&\sum_{su} m^{(1)}_{su} + m^{(2)}_{su} = m_\infty, \\
	&\sum_{su} n^{(1)}_{su} + n^{(2)}_{su} = n_\infty, \\
	&\sum_{su} h^{(1)}_{su} + h^{(2)}_{su} = \sum_{su} h^{(1)}_{su} + \tilde h^{(2)}_{su} = 0\\
	&\sum_{s} h^{(1)}_{s0} + h^{(2)}_{s0} = \sum_{s} h^{(1)}_{0s} + \tilde h^{(2)}_{s0} = 0.
\end{align}

To conclude this section, let us mention the properties of the moduli matrices that arise from $\tau_0 \to \infty$ assumption which is used to account for the quasi-static elastic response. It follows from Eqs.~\eqref{eq:tau0}, \eqref{eq:N2} and~\eqref{eq:N3} that
\begin{equation}
l^{(2)}_{0u} = m^{(2)}_{0u} = n^{(2)}_{0u} = h^{(2)}_{0u} = \tilde h^{(2)}_{0u} = 0.
\end{equation}

The obtained relations reduce the number of free parameters in the model.

\bibliographystyle{elsarticle-num}
\bibliography{refs}

\end{document}